\documentclass[pra,aps,twocolumn,showpacs]{revtex4}
\usepackage{stmaryrd}
\usepackage{amssymb}
\usepackage{amsmath}
\usepackage{epsfig}

\begin{document}

\title{The role of particle interactions in a many-body model of Feshbach
molecular formation in bosonic systems}
\author{Jing Li$^{1,2,3}$, Di-Fa Ye$^{2,4}$, Chao Ma$^{3}$, Li-Bin Fu$^{2}$,}
\author{ Jie Liu$^{1,2}$ }
\email[Email: ]{liu_jie@iapcm.ac.cn}
\affiliation{1. Center for Applied Physics and Technology, Peking University, Beijing
100084, China\\
2. Institute of Applied Physics and Computational Mathematics, P.O. Box 8009
(28), Beijing 100088, China\\
3. College of Physics and Information, Beijing Institute of Technology,
Beijing 100081, China\\
4.Graduate School, China Academy of Engineering Physics, Beijing 100088,
China}

\begin{abstract}
In this paper, we investigate the atom-molecule conversion dynamics of a
generalized many-body model that includes the atom-atom, atom-molecule, and
molecule-molecule interactions, emphasizing the efficiency of the Feshbach
molecular formation. We show that the picture of two-body molecular
production depicted by the Landau-Zener model is significantly altered: The
energy levels are dramatically distorted and the conversion efficiency is
suppressed by the particle interactions. According to the rule of constant
action and with the help of phase-space analysis, we derive an analytical
expression for the conversion efficiency in the adiabatic limit. It shows a
ceiling for the conversion efficiency when the interaction strength is
larger than a critical value. We further derive a closed equation for the
conversion efficiency with the stationary phase approximation. In the sudden
limit, the conversion efficiency is twice that predicted by the two-body
Landau-Zener formula. Our analytical formula has been confirmed by numerical
calculations.
\end{abstract}

\maketitle

\section{introduction}

The production of ultracold diatomic molecules in bosonic systems is an
exciting area of research with important applications ranging from the
production of molecular Bose-Einstein condensates (BECs)\cite{mBEC} to the
study of chemical reaction dynamics\cite{qcr}. A widely used production
technique involves the association of ultracold atoms into very weakly bound
diatomic molecules by applying a time varying magnetic field in the vicinity
of a Feshbach resonance\cite{Timmermans,rmp}. The underlying conversion
dynamics are usually described by the Landau-Zener (LZ) model\cite{lz}. In
this model, the Feshbach molecular production is discussed under a two-body
configuration where a single pair of atoms is converted to a molecule at an
avoided-crossing between atomic energy level and molecular energy level
while the molecular energy is lifted by an applied linearly sweeping
magnetic field. Thus, the molecular production efficiency is expected to be
an exponential Landau-Zener type\cite{mies,jpb}.

However, in the above two-body models\cite{mies,jpb} and their many-body
extension\cite{anglin,Bin}, interactions between particles such as the
atom-atom, atom-molecule, and molecule-molecule interactions, were not taken
into account. Recently, Santos \emph{et. al.} investigated a general
many-body two channel model in which the above particle interactions have
been taken into account in the Hamiltonian\cite{santos}. More interestingly,
the inclusion of these particle interactions brings richer dynamics to the
Josephson oscillation of the coupled atom-molecule system.

In this work, we extend to study the conversion dynamics of this general
many-body model to determine how these interactions affect molecular
conversion as the magnetic field is swept across the Feshbach resonance.

In the adiabatic limit, we cast the many-body model into an effective
classical Hamiltonian in the meanfield approximation. Such classical analog
facilitates us to analyze the conversion dynamics within the frame of
classical adiabatic theory\cite{cadia,qadia}. In this framework the
conversion dynamics are determined by two scaled parameters $U$ and $\Omega$%
, where the nonlinear interaction parameter $U$ is proportional to the
particle interactions while the coupling parameter $\Omega$ has a close
relation to resonance width. We find that the conversion efficiency can be
up to 100$\%$ when the scaled nonlinear parameter is smaller than a critical
value, \emph{i.e.}, $U/\Omega <\sqrt{2}/4$, while there is a ceiling of less
than 100$\%$ on the conversion efficiency when $U/\Omega >\sqrt{2}/4$. The
ceiling of the conversion efficiency can be calculated analytically
according to the principle of constant action. Based on this expression, we
can also obtain the scaling laws at the critical point.

In the sudden limit, we derived a closed equation for the conversion
efficiency with the stationary phase approximation. We find that the
conversion efficiency is twice of that predicted by a standard two-body
Landau-Zener formula when the sweeping rate is very large. This many body
effect is consistent with the theoretical analysis based on the renormalized
Landau-Zener formula in Ref.\cite{Bin} that has been used to explain the
experimental data\cite{hodby}.

Our paper is organized as follows: In Sec.II we introduce our model. In
Sec.III we make a comprehensive analysis of the dynamics of Feshbach
molecular formation and show how the particle interactions affect the
conversion efficiency. Sec.IV is our conclusion.

\section{Model}

A widely used molecular production technique involves the association of
ultracold atom pairs (open channel) into very weakly bound diatomic
molecules (close channel) by applying a time varying magnetic field in the
vicinity of a Feshbach resonance\cite{Timmermans,rmp}. Precisely speaking,
atoms are converted into molecules at an avoided-crossing between the atomic
energy level and the molecular energy level while the molecular energy is
lifted by an applied linearly sweeping magnetic field. The underlying
conversion dynamics are properly described by a two channel model. The
two-channel model Hamiltonian that includes the atom-atom, atom-molecule,
and molecule-molecule interactions, takes following form \cite{santos}:
\begin{eqnarray}
\hat{H} &=&\frac{u_{a}}{V}\hat{a}^{\dagger }\hat{a}^{\dagger }\hat{a}\hat{a}+%
\frac{u_{b}}{V}\hat{b}^{\dagger }\hat{b}^{\dagger }\hat{b}\hat{b}+\frac{%
u_{ab}}{V}\hat{a}^{\dagger }\hat{a}\hat{b}^{\dagger }\hat{b}  \notag \\
&&+\epsilon _{a}\hat{a}^{\dagger }\hat{a}+\epsilon _{b}\hat{b}^{\dagger }%
\hat{b}+\frac{\omega }{\sqrt{V}}(\hat{a}^{\dagger }\hat{a}^{\dagger }\hat{b}+%
\hat{b}^{\dagger }\hat{a}\hat{a}),
\end{eqnarray}%
where $\hat{a}^{\dagger }$ is the creation operator for an atomic mode while
$\hat{b}^{\dagger }$ creates a molecular mode. $\epsilon _{a}$, $\epsilon
_{b}$ are the chemical potential of the atomic mode and molecular mode,
respectively. In experiments, the external magnetic field is linearly swept $%
B(t)=\dot{B}t$, and crosses the Feshbach resonance at $B_{0}$, thus $%
2\epsilon _{a}-\epsilon _{b}=\mu _{co}[B(t)-B_{0}]$. Here, $\mu _{co}$ is
the difference between the magnetic moments of a molecule and a pair of
separated atoms. $\omega =\sqrt{4\pi \hbar ^{2}a_{bg}\Delta B\mu _{co}/m}$
denotes the amplitude for the interconversion of atoms and molecules due to
the Feshbach resonance, in which $m$ is the mass of a bosonic atom, $a_{bg}$
is the background scattering length, and $\Delta B$ is the width of the
resonance. In contrast to the early Hamiltonians\cite{anglin, Bin}, we have
included the background scattering interactions between atom-atom,
molecule-molecule, and atom-molecule, denoted by $u_{a}$, $u_{b}$, and $%
u_{ab}$, respectively. $u_{i}=4\pi \hbar ^{2}a_{i}/m_{i}$ ($i=a,b,ab$) with $%
a_{i}$ and $m_{i}$ denoting the background scattering length between
corresponding particles and their (reduced) mass, respectively. The whole
number of the system $N=N_{a}+2N_{b}$ is a constant and is very large in the
present experiments. Here, $N_{a}=\hat{a}^{\dagger }\hat{a}$ and $N_{b}=\hat{%
b}^{\dagger }\hat{b}$ are the atom number and molecule number, respectively.
We introduce the parameter $V$ to denote the quantized volume of the trapped
particles, therefore $n=N/V$ is the mean density of the initial bosonic
atoms.

Choosing the Fock states as the basis, the Schr\"{o}dinger equation is
written as
\begin{equation}
i\frac{d}{dt}|\psi \rangle =\hat{H}|\psi \rangle ,  \label{sch}
\end{equation}%
where $|\psi \rangle =\sum_{i=0}^{N/2}c_{i}|N-2i,i\rangle $, $|N-2i,i\rangle
=\frac{1}{\sqrt{(N-2i)!i!}}\left( \hat{a}^{\dagger }\hat{a}^{\dagger
}\right) ^{N/2-i}(\hat{b}^{\dagger })^{i}|0,0\rangle \quad (i=0,...,N/2)$
are Fock states, and $c_{i}$ is the probability amplitudes of the
corresponding Fock states. The normalization condition is that $%
\sum_{i}|c_{i}|^{2}=1$. The Hamitonian matrix elements are $%
H_{ij}=\left\langle N-2i,i\right\vert H\left\vert N-2j,j\right\rangle .$ For
$i=j,$ $H_{ii}=\frac{(N-2i)(N-2i-1)}{V}u_{a}+\frac{i(i-1)}{V}u_{b}+\frac{%
(N-2i)i}{V}u_{ab}+(N-2i)\varepsilon _{a}+i\varepsilon _{b};$ For $i\neq j,$ $%
H_{ij}=0$ except $H_{i,i+1}=H_{i+1,i}=\sqrt{\frac{(N-2i)(N-2i-1)(i+1)}{V}}%
\omega .$

For the simplest case of $N=2$, the above Schr\"{o}dinger equation reduces
to the following two-level system of standard Landau-Zener type,
\begin{equation}
i\frac{d}{dt}\left(
\begin{array}{c}
c_{0} \\
c_{1}%
\end{array}%
\right) =\left(
\begin{array}{ll}
2\Delta & 2\Omega \\
2\Omega & -2\Delta%
\end{array}%
\right) \left(
\begin{array}{c}
c_{0} \\
c_{1}%
\end{array}%
\right) ,
\end{equation}%
where $|c_{0}|^{2}$ and $|c_{1}|^{2}$ denote the population of atoms and
molecules, respectively. The energy bias is $\Delta =\left( 2\epsilon
_{a}-\epsilon _{b}+2nu_{a}\right) /4$ and the coupling strength is given by $%
\Omega =\omega \sqrt{n}/2$. Initially, all particles populate in the lower
level of the atomic state, \textit{i.e.}, $c_{0}=1,c_{1}=0$. When the
external magnetic field is linearly swept across the Feshbach resonance at $%
\Delta \simeq 0$, a fraction of atoms will be converted to molecules at the
avoided-crossing of energy levels. The conversion efficiency as a function
of the sweeping rate (\textit{i.e.}, $\alpha =\dot{\Delta}=\mu _{co}\dot{B}%
/4 $) and coupling strength, takes the form\cite{lz},
\begin{equation}
\chi =1-\Gamma _{lz}=1-\exp (-\frac{2\pi \Omega ^{2}}{\alpha }).  \label{lzt}
\end{equation}

It is interesting to notice that in the two-body molecular production
picture the particle interactions do not affect the conversion efficiency.
This is quite different from the many-body picture as we discuss later.

\section{Many-body effects ($N\rightarrow \infty $)}

As the total particle number $N$ increases, Eq.(\ref{sch}) is no longer
analytically solvable. In addition, the computational demand increases
dramatically as $N$ becomes very large. In the mean-field limit where $%
N\rightarrow \infty $, the quantum fluctuation is negligible. It is
appropriate to replace all the quantum operators with c-numbers, thus the
Heisenberg equations are casted into the following nonlinear Schr\"{o}dinger
equation,
\begin{equation}
i\frac{d}{dt}\left(
\begin{array}{c}
a \\
b%
\end{array}%
\right) =H\left(
\begin{array}{c}
a \\
b%
\end{array}%
\right) ,  \label{schrodinger}
\end{equation}%
where
\begin{equation}
H=\left[
\begin{array}{cc}
2U(2\left\vert b\right\vert ^{2}-\left\vert a\right\vert ^{2})+\Delta &
4\Omega a^{\ast } \\
2\Omega a & -4U(2\left\vert b\right\vert ^{2}-\left\vert a\right\vert
^{2})-2\Delta%
\end{array}%
\right] ,  \label{hamito}
\end{equation}%
with
\begin{eqnarray}
U &=&\frac{n}{4}(\frac{1}{2}u_{ab}-u_{a}-\frac{1}{4}u_{b}), \\
\Delta &=&\frac{1}{4}(2\epsilon _{a}-\epsilon _{b}+2nu_{a}-\frac{1}{2}%
nu_{b}), \\
\Omega &=&\frac{\sqrt{n}\omega }{2},
\end{eqnarray}%
and the total population is normalized to a unit, \emph{i.e.}, $\left\vert
a\right\vert ^{2}+2\left\vert b\right\vert ^{2}=1.$

We introduce the canonical transformation,
\begin{eqnarray}
S &=&\left\vert a\right\vert ^{2}-2\left\vert b\right\vert ^{2}, \\
\theta &=&2\theta _{a}-\theta _{b},
\end{eqnarray}%
where $\theta _{a}=\arg a$ is the phase of the atomic mode and $\theta
_{b}=\arg b$ is the phase of the molecular mode. The quantum system is
equivalent to the following classical Hamiltonian,
\begin{equation}
\mathcal{H}=-2US^{2}+2\Delta S+2\Omega (1+S)\sqrt{1-S}\cos \theta .
\label{H}
\end{equation}%
The new canonical variables satisfy
\begin{eqnarray}
\frac{dS}{dt} &=&-\frac{\partial \mathcal{H}}{\partial \theta }=2\Omega (1+S)%
\sqrt{1-S}sin\theta ,  \label{can1} \\
\frac{d\theta }{dt} &=&\frac{\partial \mathcal{H}}{\partial S}=-4US+2\Delta
+\Omega \frac{1-3S}{\sqrt{1-S}}cos\theta .  \label{can2}
\end{eqnarray}

Before proceeding, we would like to draw attention to some interesting
characteristics of our present Hamiltonian. Compared to the well known
nonlinear Hamiltonian that describes the tunneling dynamics of BEC atoms in
a double-well potential or between two internal states\cite{Jie}, the
nonlinearity (\emph{i.e.}, the Hamiltonian depends on the instantaneous
wavefunction as well as its conjugate) of the present Hamiltonian (\ref%
{hamito}) is much more complex. The nonlinearity stems both from the
diagonal and off-diagonal terms and is preserved even without taking into
account the particle interactions. In addition, the absence of hermicity as
well as the lack of $U(1)$ invariance of the Hamiltonian restrict the
mean-field motion to a ``tear-drop'' shaped equal-single-pair-entropy surface%
\cite{tear1,tear2}, rather than the surface of a Bloch sphere. As we will
show latter, the interplay of these new features leads to a very different
energy level structure and conversion dynamics.

In the following calculations, the coupling constant $\Omega$ is chosen as
the energy scale, thus weak nonlinearity and strong nonlinearity refer to $%
U/\Omega<<1$ and $U/\Omega>>1$, respectively.

\begin{figure}[tbp]
\centering
\rotatebox{0}{\resizebox *{8.0cm}{7.0cm}
{\includegraphics {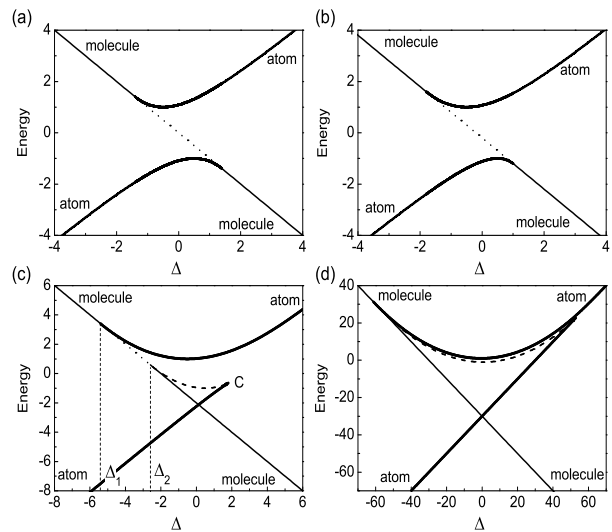}}}
\caption{Adiabatic energy levels for different nonlinear interaction
strength: (a) $U=0$, (b)$U=0.2$, (c)$U=2$, (d)$U=30$. In all cases, $%
\Omega=1 $, the solid lines represent stable eigen-states, and the dotted
lines between $\Delta_{1}=U-\protect\sqrt{2}$ and $\Delta_{2}=U+\protect%
\sqrt{2}$ correspond to unstable states. When $U>\protect\sqrt{2}/4$, a loop
structure appears at the lower energy level. The loop expands as $U$
increases.}
\label{el}
\end{figure}


\subsection{General properties}

We first show how the nonlinear interactions lead to the deformation of the
eigen-energy levels. The eigen-states of the system satisfy that
\begin{equation}
H\left(
\begin{array}{c}
a \\
b%
\end{array}%
\right) =\mu \left(
\begin{array}{c}
a \\
2b%
\end{array}%
\right) .  \label{eigen}
\end{equation}%
Solving the above nonlinear equations together with total particle
conservation condition $|a|^{2}+2|b|^{2}=1,$ we readily obtain the chemical
potential $\mu $ and the eigen-state $(a,b)$. The eigen-energies can be
derived according to the relationship $\epsilon =$ $\mu /2+\mu
|b|^{2}+\Delta |a|^{2}/2+4U|b|^{4}-2U|a|^{2}|b|^{2}.$ Their dependence on
the parameters is plotted in Fig.\ref{el} for the cases of linearity, weak
nonlinearity, and strong nonlinearity, respectively. In the linear case [$U=0
$, Fig.\ref{el}(a)], the energy levels are center symmetric. There are only
two eigen-states when $\left\vert \Delta \right\vert $ is large enough, one
for atomic mode and the other for molecular mode. When $\left\vert \Delta
\right\vert /\Omega <\sqrt{2}$ , there is an additional eigen-state with $%
S=-1$, represented by the dotted line in Fig.\ref{el}. This eigen-state is
dynamically unstable. With the appearance of nonlinear interaction, the
symmetry of the energy levels breaks down. For the weak nonlinear case $%
U/\Omega <\sqrt{2}/4$ [see Fig.\ref{el}(b)]$,$ the energy level structure is
very similar to that of the linear case except for a slight shift. However,
when $U/\Omega >\sqrt{2}/4$, a loop structure appears at the lower energy
level. The loop expands as $U$ increases, and the gap between the upper and
lower energy level becomes narrower and narrower. Such deformation of energy
levels consequently leads to very different conversion dynamics.

Consider the adiabatic evolution of the system starting from the
atomic mode at the left side of the lower energy level. When $U$ is
small, \emph{e.g.}, in Fig.\ref{el}(a), the evolution of the system
follows the solid line, converting all atoms into molecules.
However, when $U/\Omega >\sqrt{2}/4$ as in Fig.\ref{el}(c), the
system moves steadily from the left side to the critical point C.
After that, there is no way to go further except to jump to the
upper and lower levels. As that fraction of atoms tunnels to the
upper level, they are not converted into molecules. The situation
becomes even worse when $U$ is very large: the critical point is
much closer to the upper level and far away from the lower one, thus
the system will jump to the upper level more easily, see
Fig.\ref{el}(d). As a result, almost all atoms can not be converted
into molecules.

\begin{figure}[t]
\centering
\rotatebox{0}{\resizebox *{8.0cm}{6.5cm}
{\includegraphics {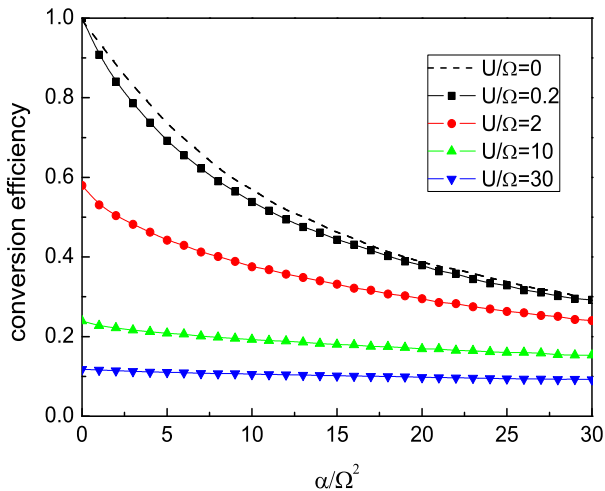}}}
\caption{(Color online) Conversion efficiency as a function of the sweeping
rate $\protect\alpha/\Omega^{2}$ for various interaction. }
\label{trans}
\end{figure}

The above simple analysis is confirmed by our numerical results, which are
plotted in Fig.\ref{trans}. In our calculations, the 4-5th Runge-Kutta
step-adaptive algorithm is used to solve the nonlinear Schr\"{o}dinger
equation (\ref{schrodinger}). The conversion efficiency as a function of the
sweeping rate $\alpha $ is plotted against the nonlinear parameters ranging
from weak nonlinearity to strong nonlinearity. Fig.\ref{trans} shows that
(1) the nonlinear interaction always suppresses the conversion from atoms to
molecules. For example, in the case of strong nonlinearity $U/\Omega =30$,
the conversion efficiency is only about 10$\%$. (2) The conversion
efficiency decreases monotonously as the sweeping rate increases. In the
following sections, we will further derive analytical formulas for the
conversation efficiency in the two limit cases, namely, the adiabatic and
sudden limit, corresponding to $\alpha /\Omega ^{2}<<1$ and $\alpha /\Omega
^{2}>>1$.

\subsection{Adiabatic limit}

\begin{figure}[t]
\centering
\rotatebox{0}{\resizebox *{8.0cm}{10.0cm}
{\includegraphics {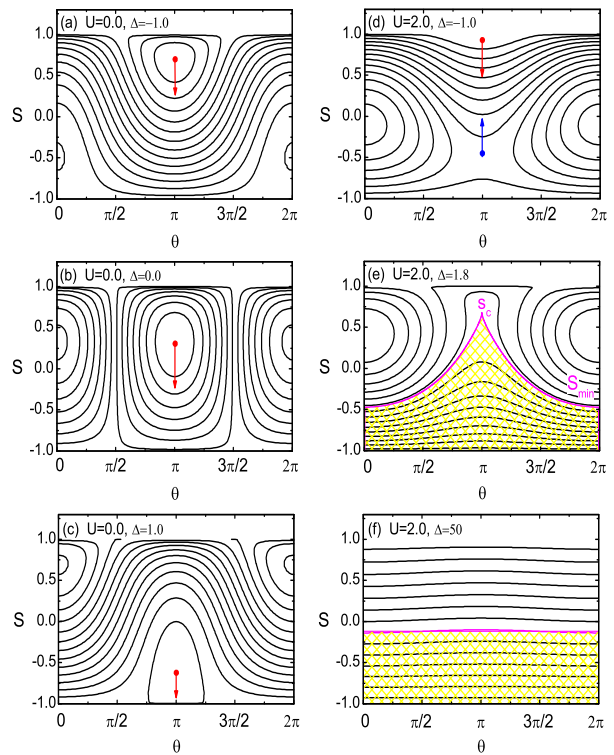}}}
\caption{(Color online) The phase space evolution as $\Delta$ changes
adiabatically. The left column and right column represent the linear and
nonlinear case, respectively. The arrows indicate the shifting direction of
the fixed points as $\Delta$ increases (see text for detail). }
\label{phase}
\end{figure}

In the adiabatic limit where the external field varies slowly compared with
the intrinsic motion of the system, the conversion dynamics are entirely
determined by the phase-space structure evolution of the classical
Hamiltonian (\ref{H}). The fixed points (\emph{i.e.}, the energy extrema of
the classical Hamiltonian) on the phase space correspond to the quantum
eigenstates. For example, in Fig.\ref{phase}(d), the elliptical point (red
circle) corresponds to the atomic mode at the left side of the lower energy
level in Fig.\ref{el}(c), and the saddle point (blue circle) corresponds to
the unstable eigenstate at the tip of the loop. According to the classical
adiabatic theory, when the energy bias $\Delta $ is modulated adiabatically,
a closed orbit in the phase-space remains closed and the action
\begin{equation}
I=\frac{1}{2\pi }\oint Sd\theta  \label{action}
\end{equation}%
stays invariant in time. The action equals the phase-space area enclosed by
the closed orbit, and is zero when the orbit shrinks to a fixed point.

For the case of $U/\Omega<\sqrt{2}/4$, the initial state [atomic mode at the
left side of the lower energy level in Fig.\ref{el}(b)] is prepared at an
elliptical point on the phase space [red circle in Fig.\ref{phase}(a)]. The
elliptical point evolves continuously from the boundary line of $S=1$ to $%
S=-1$ throughout the entire sweeping of $\Delta$. According to the rule of
constant action, it is thus expected that the system follows the elliptical
point and finally reaches $S=-1$, implying the entire conversion of atoms
into molecules, \emph{i.e.}, the conversion efficiency is $\chi=1$.

However, for the case of $U/\Omega >\sqrt{2}/4$, the elliptical point [red
circle in Fig.\ref{phase}(d)] will collide with a saddle point [blue circle
in Fig.\ref{phase}(d)] when $\Delta =\Delta _{c}$ [see Fig.\ref{phase}(e)].
After this collision, the system enters an new obit [magenta line in Fig.\ref%
{phase}(e)] with $\mathcal{H}=\mathcal{H}_{c}$, and evolves adiabatically
for $\Delta >\Delta _{c}$ according to the rule of constant action, which is
now nonzero. This obit eventually evolves into a straight line of constant $%
S $ [magenta line in Fig.\ref{phase}(f)]. With these considerations, we can
obtain the conversion efficiency in the adiabatic limit,
\begin{equation}
\chi =1-\frac{1}{2}I_{c}.
\end{equation}

To work out the explicit expression of $\chi $, we first need to determine
the critical point C. For this purpose, we notice that point C (with $\theta
=\pi $) is a double root of Eq.(\ref{can1})(\ref{can2}), thus
\begin{equation}
\left. \frac{\partial \dot{\theta}}{\partial S}\right\vert
_{S_{c}}=-4U+\Omega \frac{5-3S_{c}}{2(1-S_{c})^{3/2}}=0.  \label{1st}
\end{equation}%
The critical energy bias $\Delta _{c}$ and obit energy $\mathcal{H}_{c}$ can
be obtained through Eq.(\ref{can2}) and Eq.(\ref{H}), respectively
\begin{eqnarray}
\Delta _{c} &=&2US_{c}+\Omega \frac{1-3S_{c}}{2\sqrt{1-S_{c}}}, \\
\mathcal{H}_{c} &=&-2US_{c}{}^{2}+2\Delta _{c}S_{c}-2\Omega (1+S_{c})\sqrt{%
1-S_{c}}.
\end{eqnarray}%
Once these critical values are given, the whole orbit [magenta line in Fig.%
\ref{phase}(e)] passing through the critical point can be determined by
\begin{equation}
\cos {\theta }=\frac{\mathcal{H}_{c}+2US^{2}-2\Delta _{c}S}{2\Omega (1+S)%
\sqrt{1-S}}.  \label{rela}
\end{equation}

\begin{figure}[t]
\centering
\rotatebox{0}{\resizebox *{8.0cm}{6.0cm}
{\includegraphics {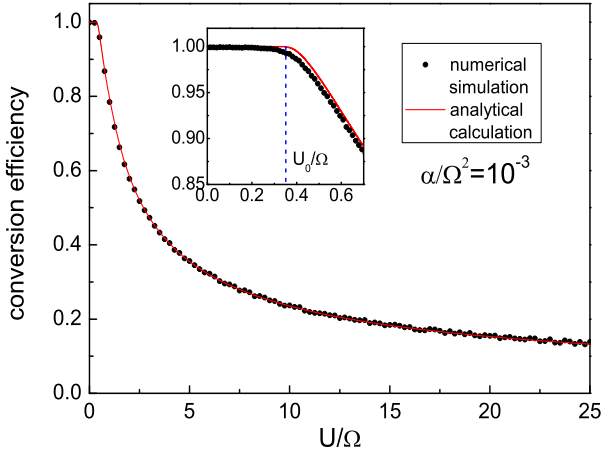}}}
\caption{(Color online) Comparison between analytical calculation (red
lines) and numerical simulation (full circles) in the adiabatic limit,
excellent agreement is clearly seen. $U_{0}/\Omega =\protect\sqrt{2}/4$.}
\label{con}
\end{figure}
We calculate the action by changing the parameter of the integration,
\begin{eqnarray}
I_{c} &=&\frac{1}{2\pi }\oint \theta dS  \notag \\
&=&\frac{1}{\pi }\int_{S_{\min }}^{S_{c}}\left[ \pi -\arccos {\left( \frac{%
\mathcal{H}_{c}+2US^{2}-2\Delta _{c}S}{2\Omega (1+S)\sqrt{1-S}}\right) }%
\right] dS  \notag \\
&&+1+S_{\min }.  \label{intgral}
\end{eqnarray}%
Here, $S_{\min }$ can be determined by Eq.(\ref{rela}) with $\theta =0$,
\emph{i.e.},
\begin{equation}
1=\frac{\mathcal{H}_{c}+2US_{\min }^{2}-2\Delta _{c}S_{\min }}{2\Omega
(1+S_{\min })\sqrt{1-S_{\min }}}.
\end{equation}

The results based on the above analytical calculation are compared with
those obtained from numerical simulation in Fig.\ref{con} and excellent
agreement is obtained. When $U/\Omega <\sqrt{2}/4$ the conversion efficiency
is always a constant, while when $U/\Omega >\sqrt{2}/4$ the conversion
efficiency decreases suddenly. This implies a phase transition at the
critical point of $U_{0}/\Omega =\sqrt{2}/4$. To confirm the occurrence of
phase transition and determine its order, we first need to obtain the
scaling law around the critical point. For this purpose, we introduce a
small variable $\delta =U/\Omega -\sqrt{2}/4$, and calculate the critical
values with perturbation theory,
\begin{eqnarray}
S_{c} &=&-1+\frac{16\sqrt{2}}{3}\delta -\frac{256}{9}\delta ^{2}+\frac{5888%
\sqrt{2}}{81}\delta ^{3}, \\
\Delta _{c}/\Omega &=&\frac{\sqrt{2}}{2}-2\delta +\frac{16\sqrt{2}}{3}\delta
^{2}-\frac{512}{27}\delta ^{3}, \\
\mathcal{H}_{c}/\Omega &=&-\frac{3\sqrt{2}}{2}+2\delta -\frac{32\sqrt{2}}{3}%
\delta ^{2}+\frac{2048}{27}\delta ^{3}.
\end{eqnarray}%
The critical orbit is determined by (let $S=-1+x$),
\begin{equation}
\cos \theta =-1+\frac{x(x-\frac{16\sqrt{2}}{3}\delta )\delta +\frac{512}{27}%
\delta ^{3}}{\sqrt{2}x}.
\end{equation}%
Here $x_{\max }\sim \frac{16\sqrt{2}}{3}\delta ,$ $x_{\min }\sim \frac{128%
\sqrt{2}}{27}\delta ^{3}$. The integration of Eq.(\ref{intgral}) results in
a power law,
\begin{equation}
\chi \sim 1-2.4\delta ^{2}=1-2.4\left( \frac{U}{\Omega }-\frac{\sqrt{2}}{4}%
\right) ^{2}, \frac{U}{\Omega }\rightarrow \frac{\sqrt{2}}{4}.
\label{power1}
\end{equation}
Clearly, both $\chi $ and its first-order derivative are continuous at the
critical point. However, its second-order derivative turns to be
discontinuous, indicating the appearance of a second order phase transition
at the critical point.

The power law at the asymptotic regime of $U/\Omega \rightarrow \infty $ can
also be obtained with the above treatment. In this case, we introduce a new
small variable $\lambda =(U/\Omega )^{-1}$ and use this small variable to
expand the critical values, $S_{c}=1-\frac{1}{4}(2\lambda )^{2/3},$ $S_{\min
}=1-\frac{9}{4}(2\lambda )^{2/3},$ $\Delta _{c}/\Omega =2\lambda ^{-1}-\frac{%
3\sqrt[3]{4}}{2}\lambda ^{-1/3}+\frac{3\sqrt[3]{2}}{4}\lambda ^{1/3},$ $%
\mathcal{H}_{c}/\Omega =2\lambda ^{-1}-3\sqrt[3]{4}\lambda ^{-1/3}+\frac{3%
\sqrt[3]{2}}{4}\lambda ^{1/3}.$ After integrating Eq.(\ref{intgral}), we
finally come to a power law of the conversion efficiency,
\begin{equation}
\chi \sim 1.2\lambda ^{2/3}=1.2\left( \frac{U}{\Omega }\right) ^{-2/3},
\frac{U}{\Omega }>>1.  \label{power2}
\end{equation}

We would like to mention that, in the nonlinear Landau-Zener model
describing the tunneling dynamics of BEC atoms in a double-well
potential or between two internal states,  the power laws at the
critical and asymptotic regime are very different\cite{Jie}. At the
critical regime, both the prefactor and exponent of the power law is
different from that in Eq.(\ref{power1}). While at the asymptotic
regime, only the prefactor is different compared with
Eq.(\ref{power2}).

\subsection{Sudden limit}

The sudden limit corresponds to nonadiabatic conversion. The conversion
efficiency is not strongly related to the structure of the energy levels. In
this limit, we can derive the analytical expression of the conversion
efficiency using the stationary phase approximation (SPA). Because of the
large sweeping rate $\alpha $, a quantum state would stay on the initial
level most of the time. Thus the amplitude $b$ in the Schr\"{o}dinger
equation (\ref{schrodinger}) remains small and $\left\vert a\right\vert \sim
1$ all the time. A perturbation treatment of the problem becomes adequate.

We begin with the variable transformation,
\begin{eqnarray}
a &=&a^{^{\prime }}\exp \left\{ -i\int_{0}^{t}\left[ \Delta +2U(2\left\vert
b\right\vert ^{2}-\left\vert a\right\vert ^{2})\right] dt\right\} , \\
b &=&b^{^{\prime }}\exp \left\{ i\int_{0}^{t}\left[ 2\Delta +4U(2\left\vert
b\right\vert ^{2}-\left\vert a\right\vert ^{2})\right] dt\right\} .
\end{eqnarray}%
As a result, the diagonal terms in the Hamiltonian are transformed away, and
the evolution equation of $b^{^{\prime }}$ becomes:

\begin{equation}
\frac{db^{^{\prime }}}{dt}=-2i\Omega (a^{^{\prime }})^{2}\exp \left\{
-i\int_{0}^{t}\left[ 4\Delta +8U(2\left\vert b\right\vert ^{2}-\left\vert
a\right\vert ^{2})\right] dt\right\} ,
\end{equation}%
here $a^{^{\prime }}\sim 1$ all the time and $\left\vert a\right\vert
^{2}+2\left\vert b\right\vert ^{2}=1$, thus

\begin{equation}
b^{^{\prime }}=-2i\Omega \int_{-\infty }^{t}dt\exp \left\{ -i\int_{0}^{t}
\left[ 4\Delta +8U(4\left\vert b\right\vert ^{2}-1)\right] dt\right\} .
\end{equation}

\begin{figure}[t]
\centering
\rotatebox{0}{\resizebox *{8.0cm}{6.5cm}
{\includegraphics {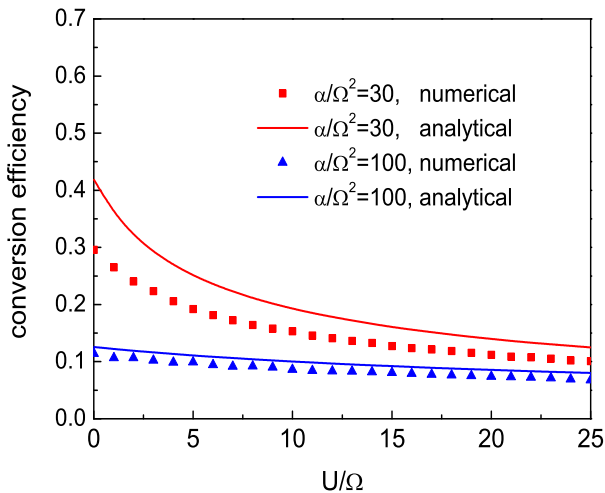}}}
\caption{(Color online) Comparison between analytical calculation (solid
lines) and numerical simulation (scatters) in the sudden limit. Good
agreement is obtained, especially for the rapidly sweeping case $\protect%
\alpha/\Omega^{2} =100$. }
\label{sudden}
\end{figure}

We need to calculate the above integral self-consistently. Due to the large $%
\alpha $, the nonlinear term in the exponent generally gives a rapid phase
oscillation, which makes the integral small. The dominant contribution comes
from the stationary point $t_{0}$ of the phase around which we have%
\begin{equation}
4\Delta +8U(4\left\vert b\right\vert ^{2}-1)=\bar{\alpha}(t-t_{0}),
\end{equation}%
with
\begin{equation}
\bar{\alpha}=4\alpha +32U\left[ \frac{d\left\vert b\right\vert ^{2}}{dt}%
\right] _{t_{0}}.  \label{alpha}
\end{equation}%
Since $\left\vert b\right\vert ^{2}=|b^{^{\prime }}|^{2}$, then we have
\begin{equation}
\left\vert b\right\vert ^{2}=\left( 2\Omega \right) ^{2}\left\vert
\int_{-\infty }^{t}dt\exp \left[ \frac{i}{2}\bar{\alpha}\left(
t-t_{0}\right) ^{2}\right] \right\vert ^{2}.
\end{equation}%
This expression can be differentiated and evaluated at time $t_{0} $. A
standard Fresnel integral with the result $[d\left\vert b\right\vert
^{2}/dt]_{t_{0}}=\left( 2\Omega \right) ^{2}\sqrt{\pi /\bar{\alpha}}$ is
obtained. Combining this with relation (\ref{alpha}), we come to a closed
equation for $\bar{\alpha}$,
\begin{equation}
\bar{\alpha}=4\alpha +32U\left( 2\Omega \right) ^{2}\sqrt{\frac{\pi }{\bar{%
\alpha}}.}  \label{close}
\end{equation}%
The conversion efficiency
\begin{equation}
\chi =2\left\vert b\right\vert _{+\infty }^{2}=\frac{16\pi \Omega ^{2}}{\bar{%
\alpha}}.
\end{equation}%
Substituting it into Eq.(\ref{close}) , we obtain the analytical expression
for the conversation efficiency,
\begin{equation}
\frac{1}{\chi }=\frac{\alpha }{4\pi \Omega ^{2}}+\frac{2U}{\pi \Omega }\sqrt{%
\chi }.  \label{rate}
\end{equation}

In Fig.\ref{sudden}, we have calculated the conversion efficiencies using
Eq.(\ref{rate}) and compared them with the numerical results obtained by
directly integrating the Schr\"{o}dinger equation (\ref{schrodinger}), where
a good agreement is shown, especially for the rapidly sweeping case, \emph{%
e.g.}, $\alpha /\Omega ^{2}=100$. In this case, the second term on the right
hand side of Eq.(\ref{rate}) can be neglected, thus we get $\chi =4\pi
\Omega ^{2}/\alpha $. This result from our many-body mean-field theory is
twice that obtained from the standard two-body Landau-Zener formula, \emph{%
i.e.}, the first order expansion of Eq.(\ref{lzt}). This many body effect is
consistent with the theoretical analysis based on a renormalized
Landau-Zener formula in Ref.\cite{Bin}.

\section{Conclusion and discussion}

In conclusion, we have both numerically and analytically investigated a
generalized many-body model that includes the atom-atom, atom-molecule, and
molecule-molecule background scattering interactions, emphasizing the
dynamics of Feshbach molecular formation. Compared to the simple two-body
molecular production picture depicted by the Landau-Zener model, the
conversion dynamics become more complicated because of the interplay of the
many-body effect and the nonlinear interaction effect.

In the adiabatic limit, the nonlinear particle interactions dominate while
in the sudden limit the many-body effect becomes significant. In the
adiabatic limit, the energy level structure is dramatically distorted due to
the particle interactions. Especially when the nonlinear parameter goes
beyond a critical value, a loop appears at the tip of the lower energy
level. This loop structure consequently leads to the break down of
adiabaticicy, and correspondingly a ceiling of less than 100$\%$ on the
conversion efficiency. The above striking phenomenon emerges when the
effective interaction parameter $U/\Omega$ is large. Because the above
parameter is proportional to the square root of the density and inversely
proportional to the square root of the resonance width, the above phenomena
are expected to be observable in experiments performed with atom clouds of
relative high density and narrow Feshbach resonance width.

In the sudden limit, we derived a closed equation for the conversion
efficiency with the stationary phase approximation. We found that the
conversion efficiency is twice that predicted by the standard two-body
Landau-Zener formula when the sweeping rate is very large. This many-body
effect is consistent with the theoretical analysis based on a renormalized
Landau-Zener formula in Ref.\cite{Bin}, that has been exploited to explain
the experimental data of JILA\cite{hodby} and agrees well.

\section{Acknowledgments}

This work is supported by National Natural Science Foundation of China
(No.10725521,10604009), the National Fundamental Research Programme of China
under Grant No. 2006CB921400, 2007CB814800.


\begin{thebibliography}{99}
\bibitem{mBEC} S. Inouye, M. R. Andrews, J. Stenger, H. -J. Miesner, D. M.
Stamper-Kum, and W. Ketterle, Nature (London) \textbf{392}, 151 (1998).

\bibitem{qcr} D. J. Heinzen and Roahn Wynar, P. D. Drummond and K. V.
Kheruntsyan, Phys. Rev. Lett. \textbf{84}, 5029 (2000).

\bibitem{Timmermans} E. Timmermans, P. Tommasini, M. Hussein, and A. Kerman,
Phys. Rep. \textbf{315}, 199 (1999).

\bibitem{rmp} Thorsten K\"ohler, Krzysztof G\'{o}ral, and Paul S. Julienne
Rev. Mod. Phys. 78, 1311 (2006).

\bibitem{lz} L. D. Landau, Phys. Z. Sowjetunion \textbf{2}, 46 (1932); G.
Zener, Proc. R. Soc. London, Ser. A \textbf{137}, 696 (1932).

\bibitem{mies} F. H. Mies, E. Tiesinga, and P. S. Julienne, Phys. Rev. A
\textbf{61}, 022721 (2000).

\bibitem{jpb} Krzysztof Goral, Thorsten Koehler, Simon A. Gardiner, Eite
Tiesinga, Paul S. Julienne, J. Phys. B \textbf{37}, 3457 (2004).

\bibitem{anglin} A. Vardi, V. A. Yurovsky, and J. R. Anglin, Phys. Rev. A
\textbf{64}, 063611 (2001).

\bibitem{Bin} Jie Liu, Bin Liu and Li-Bin Fu, arXiv:0712.4191, to appear in
Phys. Rev. A.

\bibitem{santos} G. Santos, A. Tonel, A. Foerster, and J. Links, Phys. Rev.
A \textbf{73}, 023609 (2006).

\bibitem{cadia} L.D. Landau and E.M. Lifshitz, \textit{Mechanics} (Pergamon,
Oxford, 1977).

\bibitem{qadia} Jie Liu, Biao Wu, and Qian Niu, Phys. Rev. Lett. \textbf{90}%
, 170404 (2003).

\bibitem{hodby} E. Hodby, S. T. Thompson, C. A. Regal, M. Greiner, A. C.
Wilson, D. S. Jin, E. A. Cornell, and C. E. Wieman, Phys. Rev. Lett. \textbf{%
94}, 120402 (2005).

\bibitem{Jie} Jie Liu, Li-Bin Fu, Bi-Yiao Ou, Shi-Gang Chen, and Qian Niu,
arXiv:quant-ph/0105140v1; Jie Liu, Libin Fu, Bi-Yiao Ou, Shi-Gang Chen,
Dae-Il Choi, Biao Wu and Qian Niu, Phys. Rev. A \textbf{66}, 023404 (2002).

\bibitem{tear1} I. Tikhonenkov, E. Pazy, Y. B. Band, M. Fleischhauer, and A.
Vardi, Phys. Rev. A \textbf{73}, 043605 (2006).

\bibitem{tear2} A. P. Itin and S. Watanabe, Phys. Rev. E \textbf{76}, 026218
(2007).
\end{thebibliography}
\end{document}